\documentclass[10pt,twocolumn,showpacs,amsmath,amssymb,floatfix,superscriptaddress]{revtex4-2}

\usepackage{graphicx}
\usepackage{float}
\usepackage{dcolumn}
\usepackage{bm}
\usepackage{color} 
\usepackage{txfonts}
\usepackage{microtype}
\usepackage{hyperref}
\usepackage[english]{babel}
\usepackage{slashed}
\usepackage{gensymb}
\usepackage{epsfig}
\usepackage[normalem]{ulem}
\usepackage{amsmath}
\usepackage{array}
\usepackage{booktabs}
\usepackage[english]{babel}
\usepackage{multirow}
\usepackage{CJKutf8}
\allowdisplaybreaks[4]

\begin{document}
\renewcommand{\figurename}{FIG}	

\title{Compact Spin-Polarized Positron Acceleration in Multi-Layer Microhole Array Films}

\author{Zhen-Ke Dou}
\affiliation{Ministry of Education Key Laboratory for Nonequilibrium Synthesis and Modulation of Condensed Matter, Shaanxi Province Key Laboratory of Quantum Information and Quantum Optoelectronic Devices, School of Physics, Xi'an Jiaotong University, Xi'an 710049, China}

\author{Chong Lv}
\affiliation{Department of Nuclear Physics, China Institute of Atomic Energy, P. O. Box 275(7), Beijing 102413, China}

\author{Yousef I. Salamin}
\affiliation{Department of Physics, American University of Sharjah, Sharjah, POB 26666 Sharjah,  United Arab Emirates}

\author{Nan Zhang}
\affiliation{Ministry of Education Key Laboratory for Nonequilibrium Synthesis and Modulation of Condensed Matter, Shaanxi Province Key Laboratory of Quantum Information and Quantum Optoelectronic Devices, School of Physics, Xi'an Jiaotong University, Xi'an 710049, China}
\affiliation{Guangdong Provincial Key Laboratory of Semiconductor Optoelectronic Materials and Intelligent Photonic Systems, Shenzhen 518055, China}

\author{Feng Wan}\email{wanfeng@xjtu.edu.cn}
\affiliation{Ministry of Education Key Laboratory for Nonequilibrium Synthesis and Modulation of Condensed Matter, Shaanxi Province Key Laboratory of Quantum Information and Quantum Optoelectronic Devices, School of Physics, Xi'an Jiaotong University, Xi'an 710049, China}

\author{Zhong-Feng Xu}
\affiliation{Ministry of Education Key Laboratory for Nonequilibrium Synthesis and Modulation of Condensed Matter, Shaanxi Province Key Laboratory of Quantum Information and Quantum Optoelectronic Devices, School of Physics, Xi'an Jiaotong University, Xi'an 710049, China}

\author{Jian-Xing Li}\email{jianxing@xjtu.edu.cn}
\affiliation{Ministry of Education Key Laboratory for Nonequilibrium Synthesis and Modulation of Condensed Matter, Shaanxi Province Key Laboratory of Quantum Information and Quantum Optoelectronic Devices, School of Physics, Xi'an Jiaotong University, Xi'an 710049, China}	
\affiliation{Department of Nuclear Physics, China Institute of Atomic Energy, P. O. Box 275(7), Beijing 102413, China}

	
\begin{abstract}

Compact spin-polarized positron accelerators play a major role in promoting significant positron application research, which typically require high acceleration gradients and polarization degree, both of which, however, are still great challenging. Here, we put forward a novel spin-polarized positron acceleration method which employs an ultrarelativistic high-density electron beam passing through any hole of multi-layer microhole array films  to excite strong electrostatic and transition radiation fields. Positrons in the polarized electron-positron pair plasma, filled in the front of the multi-layer films, can be captured, accelerated, and focused by the electrostatic and transition radiation fields, while maintaining high polarization of above $90\%$ and high acceleration gradient of about ${\rm TeV/m}$. Multi-layer design allows for capturing more positrons and achieving cascade acceleration. Our method offers a promising solution for accelerator miniaturization, positron injection, and polarization maintaining, and also can be used to accelerate  other charged particles.

\end{abstract}
	
\maketitle

Spin-polarized positron beams are widely used in medical diagnosis \cite{Raichle1985}, materials physics \cite{Gidley1982, Christoph2016, Selim2021}, laboratory astrophysics \cite{Chenhui2023}, and high-energy physics \cite{Musumeci2022, Moortgat2008, Ari2016, Duan2019, Chaikovska2022}, etc. Such as studying $\gamma$-ray bursts \cite{Sarri2015, Warwick2017}, measuring the proton elastic form factor ratio \cite{Bernauer2018, Boyko2023} and nucleon electromagnetic structure \cite{Afanasev2019}, testing the Standard Model \cite{Androic2013}, and searching for new physics beyond the Standard Model \cite{Moortgat2008}. In these applications, there is high demand for positron beam density (exceeding $10^{16}\:\rm{cm^{-3}}$), energy (hundreds of ${\rm MeV}$ to ${\rm TeV}$), and polarization degree ($>30\:\%$) \cite{Chenhui2023, Moortgat2008, Scott2011, Vauth2016}. Currently, their generation mainly relies on large-scale conventional accelerators with acceleration gradients below ${\rm 100\:MeV/m}$ \cite{Gschwendtner2019}. A compact spin-polarized positron accelerator would play a major role in promoting research into the above, and possibly other, applications \cite{Ruffini2010, Harding2006}.

Alternative mechanisms currently being explored towards reaching that goal include dielectric laser acceleration \cite{Breuer2013, Peralta2013, England2014, Wei2017} and solid-state wakefield acceleration using crystals\cite{Tajima1987, Chen1997, Sahai2019} or carbon nanotube arrays \cite{Gilljohann2023, Bonatto2023, MartínLuna2023}, which can achieve gradients of tens of ${\rm GeV/m}$ and even ${\rm TeV/m}$. However, energy gain in the former is limited due to the short laser-particle interaction length \cite{Wei2017} and the latter requires ultra-high power x-ray lasers to excite crystals \cite{Bonatto2023}. Meanwhile, significant progress has been made over the past decade in plasma wakefield acceleration (PWFA), which can achieve gradients of hundreds of ${\rm GeV/m}$ \cite{Chen1985, Litos2014, Tajima1979, Esarey2009}. PWFA is well-suited for trapping \cite{Kalmykov2009} and accelerating electrons \cite{Lu2007}, with gradients sustained over meters \cite{Blumenfeld2007}, high energy transfer efficiency \cite{Litos2014}, and low energy spread in the bubble regime \cite{Wu2019, Pompili2021}. Effective trapping and acceleration of positrons, however, is limited by defocusing due to the transverse field in the bubble.

To ameliorate these limitations, various methods have been proposed \cite{Cao2024}, such as using a long positron beam to achieve energy transfer from head-to-tail in the self-loaded plasma wakefields \cite{Blue2003, Corde2015}. Hollow electron beams \cite{Jain2015}, Laguerre-Gaussian laser pulses \cite{Vieira2014}, finite radius plasma columns \cite{Diederichs2019, Diederichs2020}, plasma channel \cite{Liu2023}, and two-column plasma structures \cite{Reichwein2022} are used to control wakefield structures for simultaneous accelerating and focusing positrons in the plasma bubble. Another attractive method involves using a hollow plasma channel, for which the longitudinal field is uniform in the transverse plane, and the transverse focusing fields vanish inside the channel, ensuring beam emittance preservation during acceleration \cite{Yi2014, Gessner2016, LiY2019, Silva2021, Zhou2022}. The beam-breakup instability can be avoided by utilizing an asymmetric electron beam in hollow plasma channel \cite{Zhou2021}. Nevertheless, all of these methods require precise injection of the high-energy positrons, with their beam polarization properties often overlooked. Meanwhile, density of the accelerated positron beam seems very limited (less than $10^{18}\:\rm{cm^{-3}}$). A single-stage mechanism for polarized positron production, injection, and acceleration, has also been proposed through the interaction of a seed electron beam with a strong laser \cite{Liu2022, Martinez2023, Vranic2018}. In this case, polarized positrons can be generated via nonlinear Compton scattering and nonlinear Breit-Wheeler process \cite{FEDOTOV2023, DiPiazza2012, Gonoskov2022, LiYanFei2020, XueKun2020, WanFeng2020, ChenYueYue2019}. However, acceleration in a single stage requires precise spatiotemporal synchronization, which presents experimental difficulties, and faces large depolarization challenges due to spin precession in the fields with complicate structures \cite{Liu2022}.

Another positron acceleration scheme relies on mid-infrared \cite{si2023} and terahertz \cite{Zhao2023, XuZhangli2023} radiation, both of which still require precise injection and overlook the polarization properties. Moreover, circularly polarized $\gamma-$photons \cite{Alexander2008, Omori2006} or polarized electron beams \cite{Abbott2016} hitting high-Z targets can generate polarized positrons via the Bethe-Heitler process \cite{Myatt2009}. The positrons can be accelerated by coherent transition radiation \cite{Xu2020} or sheath fields \cite{Yan2017}. However, the acceleration gradients are only ${\rm GeV/m}$, and the positron beam density is less than $10^{15}\:{\rm cm^{-3}}$. Thus, realization of a compact high-density polarized positron accelerator is still a great challenge.

\begin{figure}[t] 
	\begin{center}
		\includegraphics[width=\linewidth]{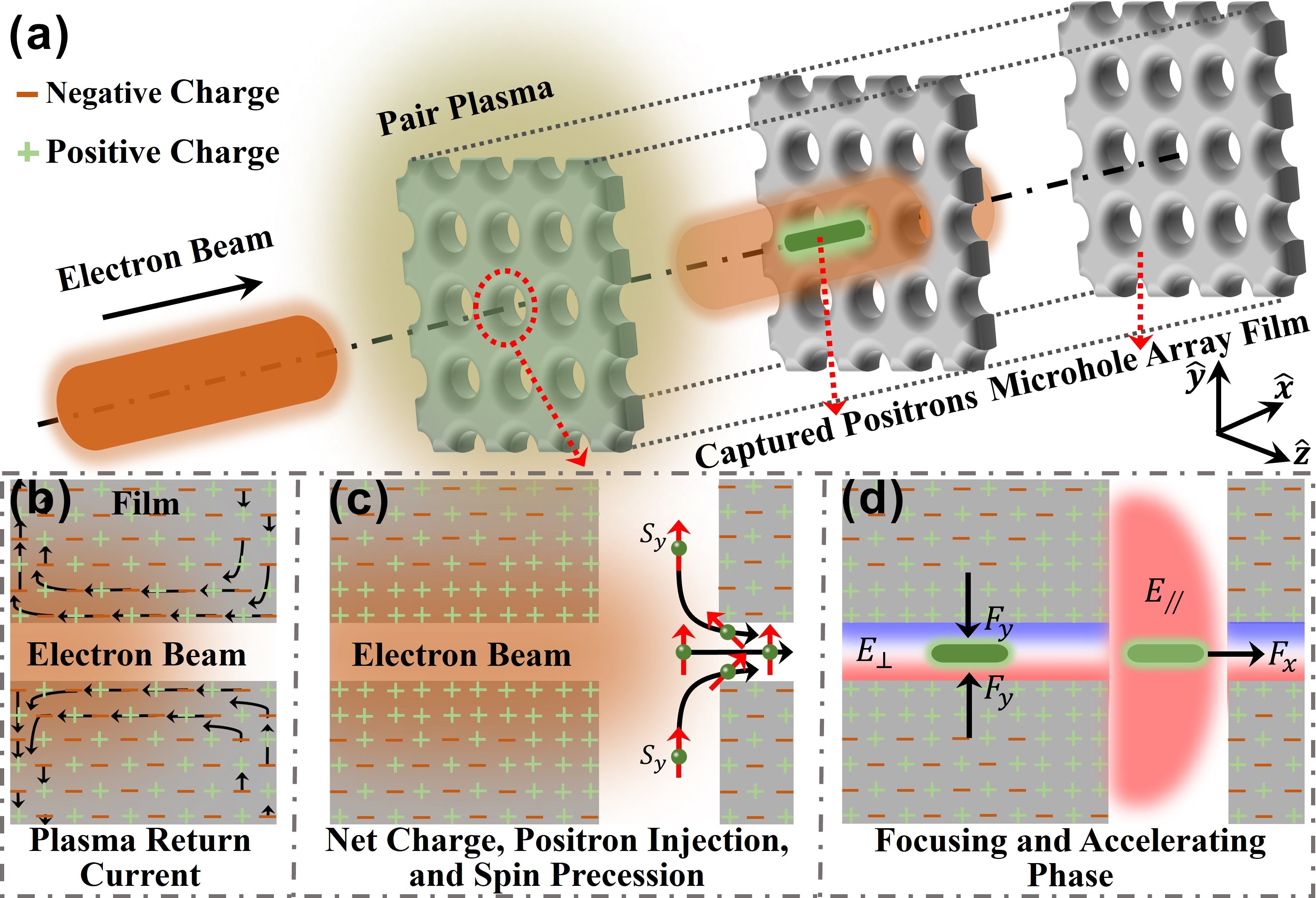}
		\caption{(a) Electron beam passing through any hole of the multi-layer microhole array films, with front (one layer shown) filled with spin-polarized electron-positron pair plasma. Positrons in the pair plasma can be effectively captured, accelerated, and focused, while maintaining high polarization. (b) Plasma return current. Black arrows represent the direction of film's electron motion. (c) Net charge, positron injection, and spin precession. Green balls represent the captured positrons. Black-solid lines with arrow and red arrows depict positron trajectory and spin component $S_y$, respectively. (d) Focusing and accelerating phase: $E_\perp$ and $E_\varparallel$ (blue-red gradients) within and behind the hole focus and accelerate the positron beam. Forces $F_y$ and $F_x$ acting on a positron are represented by black arrows.}\label{fig1}
	\end{center}
\end{figure}

In this Letter, we put forward a novel polarized positron acceleration method, which uses an ultrarelativistic high-density electron beam passing through any hole of multi-layer microhole array films; see Fig.$\:$1(a). The electron beam ionizes the film around the hole creating plasma and inducing plasma return current in the film \cite{Miller2012}; see Fig.$\:$1(b). This current causes net charge accumulation on the film surface; see Fig.$\:$1(c). The net charge excites the transverse and longitudinal electrostatic fields ($E_{y, {\rm sta.}}$ and $E_{x, {\rm sta.}}$) within and behind the hole, which generates focusing field $E_\perp\varpropto E_{y, {\rm sta.}}$ and acceleration field $E_\varparallel\varpropto E_{x, {\rm sta.}}$ for positrons; see Fig.$\:$1(d). When the electron beam crosses the plasma-vacuum boundary (film's rear surface), it excites the transition radiation field $E_{\rm rad.}$ behind the hole \cite{Bolotovskii2009}. The longitudinal component $E_{x, {\rm rad.}}$ can significantly enhance the acceleration field via $E_ \varparallel=E_{x, {\rm sta.}}+E_{x, {\rm rad.}}$. The electron-positron pair plasma (which can be generated by radiation sources \cite{Greaves1995}, nuclear reactions \cite{Pedersen2012, Hugenschmidt2012, Stenson2018, Stoneking2020}, particle accelerators \cite{Arrowsmith2021, Dumas2009}, Bethe-Heitler processes \cite{Chenhui2023, Sarri2015, KIM2024}, and Breit–Wheeler processes \cite{XueKun2023, Song2022, zhuxinglong2016, LuoWen2018, HeYutong2021, GuYanJun2019, GuYanJun2018}) fills the front of the multi-layer films; see Fig.$\:$1(a). Positrons within the pair plasma are attracted (electrons are repelled) towards the central axis ($y=z=0$) by the electron beam and then accelerated in the $+\hat{x}$ direction by $E_\varparallel$; see Fig.$\:$1(c). Subsequently, they enter the focusing and accelerating phase, forming a high-density polarized positron beam, which can be focused and accelerated to high energies; see Fig.$\:$1(d). In our method, the acceleration gradient for the positron beam can reach about the order of ${\rm TeV/m}$, which holds significant potential for accelerator miniaturization and cost-effectiveness. Depolarization caused by spin precession [see Fig.$\:$1(c)] is small because most positrons near the central axis experience weak magnetic fields. Compared to slits and polygonal holes, circular holes have a better quality factor for focusing positron beams \cite{Lukin2023, Reiserer2015}. Due to the strong self-generated field (ionized film) of the electron beam, the material of the film can be metal, plastic, etc. Our method is robust with respect to the multi-layer films, electron beam, and pair plasma parameters; see Fig.$\:$S1 in \cite{SM}. These advantages can lower the level of difficulty in running the experiment. Moreover, our method has the potential to accelerate electrons using positively charged particle beams; see Figs.$\:$S2 and S3 in \cite{SM}.

We perform spin-resolved quantum electrodynamics particle-in-cell simulations, by using the spin-resolved SLIPs code \cite{Wan2023, XueKun2023} to illustrate the dynamics of positrons; see simulation details in \cite{SM}. We also use cylindrical coordinate system ($r,\phi,x$), with $r=\sqrt{y^2+z^2}$, $\phi=\arctan(z/y)$, and $x$ being the radial, azimuth, and vertical directions, respectively. The plasma critical density $n_c=\omega^2m_e\epsilon_0/e^2$, period $T_0=\lambda/c$, $E_0=m_ec\omega/e$, and $B_0=m_e\omega/e$ are used for the normalization, where $\epsilon_0$ is the vacuum permittivity constant, $m_e$ is electron mass, $e$ is elementary charge, $\lambda=1\:{\rm \mu m}$ is reference wavelength, and $\omega=2\pi c/\lambda$ is reference frequency, respectively. 

The typical positron acceleration results are shown in Fig.$\:$2, with the following simulation parameters. A moving window along $+x$-direction is used, with a simulation box covering $0\leq x \leq6\:{\rm \mu m}$, $-3\:{\rm \mu m}\leq y \leq 3\:{\rm \mu m}$, and $-3\:{\rm \mu m}\leq z \leq 3\:{\rm \mu m}$ and divided into $600 \times 600 \times 600$ cells. Electron beam number density profile is $n_b=n_{b0}\exp(-r^2/2\sigma_r^2)\exp(-(x-x_0-v_bt)^2/2\sigma_x^2)$, and truncated at $n_b=0.02\:n_c$. Here, $\sigma_x = \sigma_r=0.85\:{\rm \mu m}$ are the longitudinal and transverse rms-length (the radius of the electron beam is $R_b=2\:{\rm \mu m}$), $v_b=c\sqrt{1-1/\gamma_b^2}$ is the beam velocity, $\gamma_b=20000$ is the beam relativistic factor, $x_0=4\:{\rm \mu m}$ is the beam center position at $t=0$ (when the electron beam contact with the front surface of first layer film at $x=0$), and $n_{b0}=0.4\:n_c$ is the peak number density, respectively. The electron beam parameters are feasible in e.g., PWFA \cite{Gonsalves2019, ZhuXingLong2020}, the FACET-II facilities \cite{Yakimenko2019}, and plasma lens \cite{Doss2019}. We consider a partially ionized multi-layer polystyrene films composed of $e^-$, $H^+$, and $C^{6+}$ ions, with number densities $n_e(e^-)=30\:n_c$ and $n_p(H^+)=n_C(C^{6+})=n_e(e^-)/7$; see Fig.$\:$S4 in \cite{SM}. The film thickness is $L=1\:{\rm \mu m}$, the layer spacing is $D=1.5\:{\rm \mu m}$, and the duty ratio is $D_r=L/(L+D)=40\:\%$ in $x$-direction, respectively. The hole radius is $R_h=0.5\:{\rm \mu m}$ ($R_h<R_b$) and the distance between the centers of adjacent holes in $y$ and $z$ directions is $R_c\geq R_b + R_h$ (to achieve the maximum acceleration field under these parameter conditions). Polarized pair plasma, generated by ultraintense laser pulses \cite{Yoon2019, Yoon2021, Danson2019, ELI-beamlines, ELI_np, SULF, ECELS} irradiating a solid target \cite{XueKun2023, Song2022}, has a uniform number density distributions with $0.01\:n_c$ in the region of $1\:{\rm \mu m}\leq x \leq 20\:{\rm \mu m}$, $-3\:{\rm \mu m}\leq y\leq 3\:{\rm \mu m}$, and $-3\:{\rm \mu m}\leq z\leq 3\:{\rm \mu m}$. Each cell includes $10$ macroelectrons (electron beam), $10$ macroelectrons, $5$ macroprotons, and $5$ macrocarbon ions (film), and $10$ macropositrons and $10$ macroelectrons (pair plasma).

\begin{figure}[t] 
	\begin{center}
		\includegraphics[width=\linewidth]{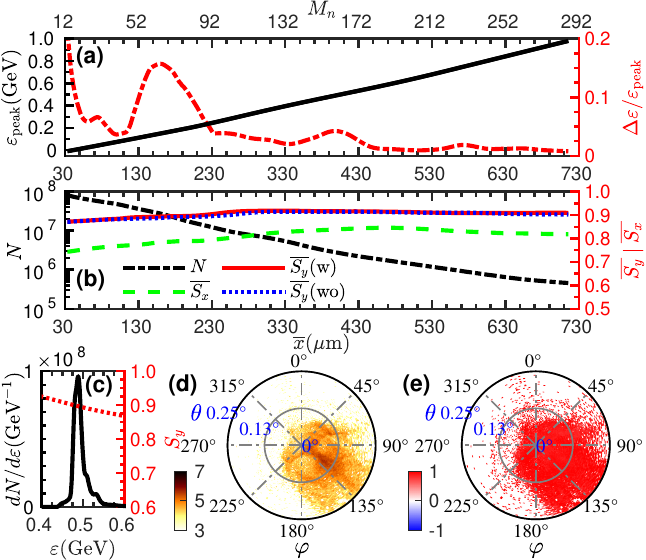}
		\caption{(a) Evolution of the peak energy $\varepsilon_{\rm peak}$ (black solid) and relative energy spread $\Delta\varepsilon/\varepsilon_{\rm peak}$ (red dashed-dotted) of the positron beam as a function of the mean longitudinal position $\overline{x}$ of positrons and layer number $M_n$. (b) Total number $N$ (black dashed-dotted), average transverse spin polarization degree $\overline{S_y}$ (red solid: with radiation, and blue dotted: without radiation, artificially neglecting the positron radiation), and average longitudinal spin polarization degree $\overline{S_x}$ (green dashed) of positron beams vs $\overline{x}$. (c) Energy spectra $dN/d\varepsilon$ (black solid) and transverse spin component $S_y$ (red dotted) vs energy $\varepsilon$ of positrons at $\overline{x}=411\:{\rm \mu m}$ ($t = 413\:T_{\rm0}$). (d),(e) ${\rm log_{10}}(d^2N/d\theta d\varphi)$ and $S_y$ with respect to the polar angle $\theta$ and azimuth angle $\phi$ at $\overline{x}=411\:{\rm \mu m}$.}\label{fig2}
	\end{center}
\end{figure}      

\begin{figure}[t] 
	\begin{center}
		\includegraphics[width=\linewidth]{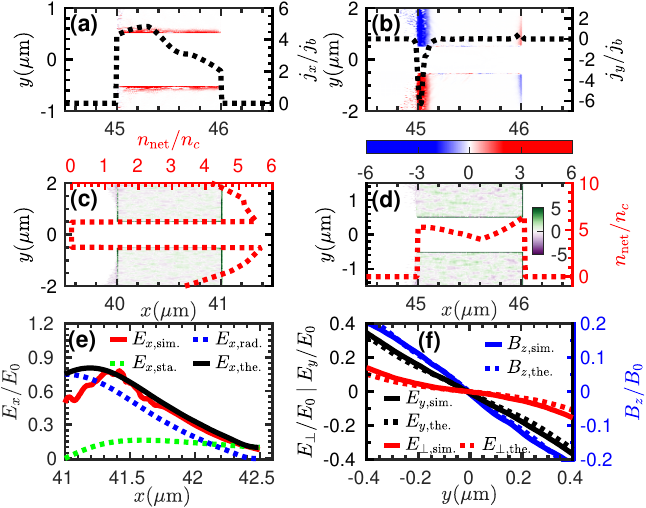}
		\caption{(a),(b) Current density $j_x$ and $j_y$ distributions (blue-red gradients) at $t=47\:T_{\rm0}$. The black-dotted lines represent $j_x$ at $y=0.52\:{\rm \mu m}$ and $j_y$ at $y=1\:{\rm \mu m}$, vs $x$. (c),(d) Distributions of $n_{\rm net}$ (purple-green gradients) at $t=43\:T_{\rm0}$ and $t=47\:T_{\rm0}$, respectively. Red-dotted lines represent $n_{\rm net}$, at $x=41\:{\rm \mu m}$ vs $y$, and at $y=0.5\:{\rm \mu m}$ vs $x$. (e) Acceleration field $E_x$ on the central axis vs $x$. Red-solid line represents simulation result, while the green-dotted, blue-dotted, and black-solid lines represent theoretical analysis results of electrostatic field, transition radiation field, and $E_{x,{\rm the.}}=E_{x,{\rm sta.}}+E_{x,{\rm rad.}}$, respectively. (f) Electric field $E_y$, magnetic field $B_z$, and focusing field $E_\perp=E_y-cB_z$, at $x=45.9\:{\rm \mu m}$, vs $y$. Black-solid, blue-solid, and red-solid lines represent the simulation results, while the black-dotted, blue-dotted, and red-dotted lines represent the theoretical analysis results, respectively. 
		} \label{fig3}
	\end{center}
\end{figure}   

The multi-layer design allows for cascade acceleration of the positron beam, increasing its peak energy by $\varepsilon_g\sim3.6\:{\rm MeV}$ per layer, and determining the final peak energy as $\varepsilon_{\rm peak}=\varepsilon_g M_n$; see Fig.$\:$2(a). As the positrons accelerate, $\Delta\varepsilon/\varepsilon_{\rm peak}$ decreases oscillatingly to below $2.5\:\%$ in the later stages of acceleration; see Fig.$\:$2(a). The main cause of oscillation is the weak defocusing field in the acceleration phase, which leads to the escape of some positrons. This, in turn, alters the shape of the positron beam energy spectrum. The escape of positrons lead to a gradual decrease in $N$, approaching $4.6\times10^{5}$ (density $>10^{20}\:{\rm cm^{-3}}$) at $\overline{x}=715\:{\rm \mu m}$; see Figs.$\:$2(b). Due to the weak magnetic fields experienced by most positrons, the depolarization caused by precession is minimal; see Fig.$\:$4. Moreover, the effect of radiation and spin flipping can be negligible; see Figs.$\:$2(b) and S5 in \cite{SM}. Therefore, the positron beam has transverse (longitudinal) spin polarization if the pair plasma initially has transverse (longitudinal) polarization, with average spin polarization degree $\overline{S_y}=|\sum_i^{N} S_y/N|$ ( $\overline{S_x}=|\sum_i^{N} S_x/N|$) exceeding $90\%$ ($80\%$), and it remains virtually unchanged during the acceleration process; see Fig.$\:$2(b). In the case of longitudinal spin, due to positron's spin direction being perpendicular to the circular magnetic field everywhere, spin precession has a greater impact, resulting in a lower $\overline{S_x}$ compared to $\overline{S_y}$; see Fig.$\:$2(b). For the electron beam, due to radiation, it has the potential to be realized beam-self polarization; see Fig.$\:$S6 in \cite{SM}. The energy spectrum shows a single peak and $S_y$ decreases gradually with the increase in energy, surpassing $0.85$ within the full width at half maximum; see Fig.$\:$2(c). The positron beam is well-collimated, majority of positrons are located within $\theta < 0.25\degree$, and exhibit a uniform $S_y$ distribution with $\theta$ and $\phi$; see Figs.$\:$2(d) and 2(e). This result holds great potential for studies of $\gamma$-ray bursts \cite{Sarri2015, Warwick2017}, pulsar magnetospheres \cite{Brambilla2018}, and detection of dark matter particles \cite{Boehm2004}. Moreover, the externally injected positron and proton beams can still be accelerated well; see Figs.$\:$S7 and S8 in \cite{SM}. The maximum energy conversion efficiency can reach up to $25.5\%$, and the emittance stabilizes in the later stages of acceleration; see Fig.$\:$S9 in \cite{SM}. The annihilation of electron-positron can be negligible; see Fig.$\:$S10 in \cite{SM}. 

The generation mechanism of acceleration and focusing field in $x$-$y$ plane are shown in Fig.$\:3$. The return current has a high-density $j_{\rm ret.}\sim \left |j_{b}\right | L/2l_s{\exp}(1)\approx 6|j_b|$ \cite{Miller2012, Jiang2023}; see Figs.3(a) and 3(b), where $l_s=c/\omega_{\rm p}=0.029\:{\rm \mu m}$, $\omega_{\rm p}=\sqrt{n_ee^2/m_{e}\epsilon_0}$, and $j_b=-e v_b n_b\approx-2.14\times 10^{16}\:{\rm A/m^2}$ are the skin length, plasma frequency, and electron beam current density, respectively. This return current accumulates net charge on the surface of the film; see Figs.$\:$3(c) and 3(d), whose magnitude is proportional to the integral of $j_{\rm ret.}$ over duration of the electron beam’s action on the film, namely $n_{\rm net}\sim (j_{\rm ret.}\times D_n/v_b)/e\approx4.5\:n_c$, where $D_n$ is the distance from the front of the electron beam to the rear surface of the $n$th layer film. Behind the hole ($41\:{\rm \mu m}\leq x\leq42.5\:{\rm \mu m}$), the longitudinal acceleration field is composed of an electrostatic part, excited by the net charge, and a transition radiation part, excited by the electron beam. Distribution of the net charge can be approximately modeled by that of a uniformly charged disk of thickness $l_s$, inner radius $R_1\approx R_h$, and outer radius $R_2\approx1.5\:{\rm \mu m}$; see Fig.$\:$3(c). The electrostatic field strength at any point on the central axis can be approximated by $E_{x,{\rm sta.}}=\frac{(en_{\rm net}l_s)(x-x_n)}{2\epsilon_0}[\frac{1}{(R_1^2+(x-x_n)^2)^{1/2}}-\frac{1}{(R_2^2+(x-x_n)^2)^{1/2}}]$ of an ideal disk of charge; see Fig.$\:$3(e), where $x_{n}$ is rear surface position of the $n$th layer film. The transition radiation field can be found from $E_{x,{\rm rad.}}=\sum_{1}^{N_e^n}\frac{1}{4\pi\epsilon_0}\frac{e}{c\pi R}\frac{{\rm sin} \omega_{\rm p}(t-R/c)}{t-R/c}\frac{v_e/c{\rm sin}^2\theta}{1-(v_e/c)^2{\rm cos}^2\theta}$ \cite{Bolotovskii2009}; see Fig.$\:$3(e), where $N_e^n\approx1.5\times10^9$ is the total number of electrons in the electron beam, extending from $x_n$ to $x_n+D$, $v_e\approx v_b$ is the electron velocity, and $R=\sqrt{(x-x_n)^2+y^2+z^2}$ is distance between the observation point and rear surface position of the $n$th layer film, respectively. The total longitudinal acceleration field is $E_{x,{\rm the.}}=E_{x,{\rm sta.}}+E_{x,{\rm rad.}}$, in agreement with the simulation result $E_{x,{\rm sim.}}$; see Fig.$\:$3(e). 

\begin{figure}[t] 
	\begin{center}
		\includegraphics[width=\linewidth]{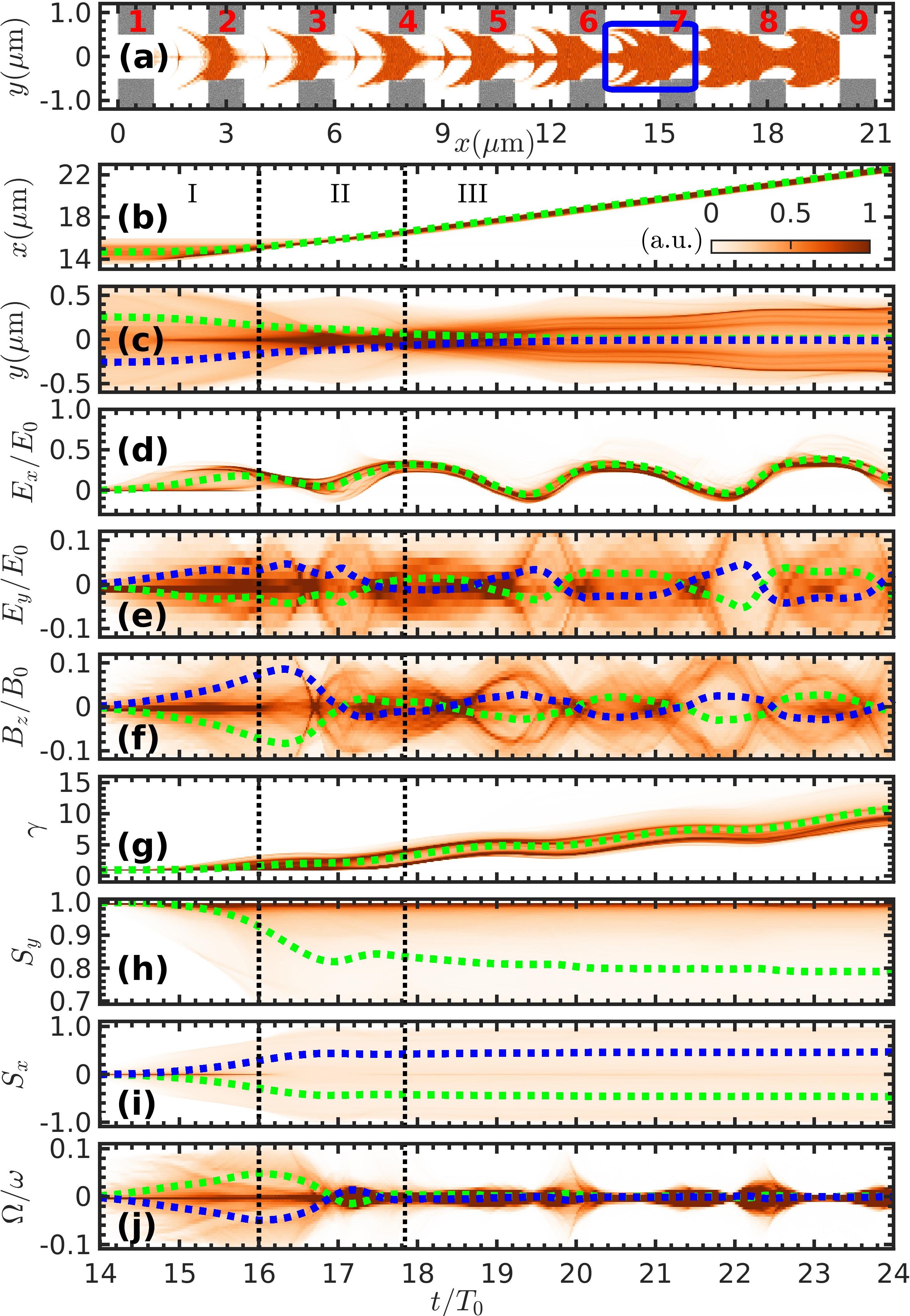}
		\caption{(a) Multi-layer films (gray) and positions of captured positrons at $t=0$ (orange). The red numbers represent the sequence number of each layer film. (b)-(j) $dN/dx$, $dN/dy$, $dN/dE_x$, $dN/dE_y$, $dN/dB_z$, $dN/d\gamma$, $dN/dS_y$, $dN/dS_ x$, and $dN/d\Omega$ of the positrons (white-orange gradients) injected from blue rectangle in (a) with respect to time $t$, where $N$ is the positron number. (c), (e), (f), (i), and (j) Green-dotted and blue-dotted lines show the corresponding average values, with initial positron transverse positions $y>0$ and $y<0$. (b), (d), (g), and (h) Green-dotted line shows the corresponding average values. 
		}\label{fig4}
	\end{center}
\end{figure}  

Within the hole ($45\:{\rm \mu m}\leq x\leq46\:{\rm \mu m}$), the transverse field is composed of the transverse electrostatic field, the self-generated field of the electron and positron beams, and the magnetic field generated by the return current. Distribution of the net charge can be approximately modeled by that of a uniformly charged column shell of length $L$, inner radius $R_3\approx R_h$, and outer radius $R_4\approx R_h + 1.5l_s$; see Fig.$\:$3(d). The transverse electrostatic field excited by this column shell can be approximated by $E_{r,\rm {sta.}}=\frac{2en_{\rm net}}{\pi \epsilon_0}\int_{0}^{\infty}[R_3K_1(kR_3)-R_4K_1(kR_4)][-I_1(kr){\rm cos}(kx)]{\rm sin}(k\frac{L}{2})\frac{dk}{k},(k=1,2,3\cdots$), where $K_1$ and $I_1$ are modified Bessel functions. Moreover, initial self-generated field of the electron beam has the components: $E_r^b=\frac{1}{4\pi\epsilon_0}\frac{-2eN_e}{\sqrt{2\pi}\sigma_x}\frac{1}{r}(1-{\exp}(\frac{-r^2}{2\sigma_r^2})){\exp}(-(x-x_0-v_bt)^2/2\sigma_x^2)$, $E_x^b=0$, and $B_\phi^b=\sqrt{\mu_0\epsilon_0}\frac{v_b}{c}E_r^b$ \cite{Sampath2020, Sampath2021}, where $N_e\approx3.8\times10^9$ is the total number of electrons in the electron beam. Similarly, the field generated by the positron beam is given by $E_r^{b'}=\frac{1}{4\pi\epsilon_0}\frac{2eN_p}{\sqrt{2\pi}\sigma'_x}\frac{1}{r}(1-{\exp}(\frac{-r^4}{{\sigma'_r}^4})){\exp}(-(x-vt)^4/{\sigma'_x}^4)$, $E_x^{b'}=0$, and $B_\phi^{b'}=\sqrt{\mu_0\epsilon_0}\frac{v}{c}E_r^{b'}$, where $N_p\approx6.8\times10^7$ is the total number of positrons in the positron beam, $\sigma'_x\approx0.3\:{\rm \mu m}$, and $\sigma'_r\approx0.2\:{\rm \mu m}$, respectively. In the $x$-$y$ plane, the transverse electric field components are: $E_{y,{\rm sta.}}=E_{r,{\rm sta.}}{\rm cos}\phi$, $E_y^b=E_r^b{\rm cos}\phi$, and $E_y^{b'}=E_r^{b'}{\rm cos}\phi$. The total transverse electric field is $E_{y,{\rm the.}}=E_{y,{\rm sta.}}+E_y^b+E_y^{b'}$, consistent with $E_{y,\rm{sim.}}$; see Fig.$\:$3(f). Furthermore, the plasma return current can be approximated as a finite-length current-carrying cylindrical surface with a radius $R_5\approx R_h$; see Fig.$\:$3(a). The magnetic field is, therefore, given by $B_{r,{\rm ret.}}\approx\frac{\mu_0\overline{j_{\rm ret.}}}{4\pi r}(1+\frac{r^2-R_5^2}{\sqrt{(R_5^2-r^2)^2}})$. Inside the cylindrical surface ($r<R_h$), $B_{r,{\rm ret.}}=0$. In the $x$-$y$ plane, the transverse magnetic field is $B_z^b=B_\phi^b{\rm cos}\phi$ and $B_z^{b'}=B_\phi^{b'}{\rm cos}\phi$. The total transverse magnetic field is $B_{z,{\rm the.}}=B_z^b+B_z^{b'}$, in agreement with $B_{z,\rm{sim.}}$; see Fig.$\:$3(f). Hence, the total transverse focusing field is $E_{\perp,{\rm {\rm the.}}}=E_{y,{\rm {\rm the.}}}-cB_{z,{\rm the.}}$, in good agreement with $E_{\perp,\rm{sim.}}$; see Fig.$\:$3(f). Because $E_y^b-cB_z^b=0$ and $E_y^{b'}-cB_z^{b'}=0$, the focusing force is only provided by $E_{y,\rm{sta.}}$. Moreover, the beam loading effect makes $E_\varparallel$ more gentle, which can improve the quality of the positron beam \cite{Silva2021}, while having little effect on $E_\perp$; see Figs.$\:$S11, S12, and S13.

Due to the cylindrical symmetry of our method, for simplicity, we use two-dimensional particle tracking simulation in Fig.$\:4$ to illustrate the capture and spin evolution of the positrons. Other parameters are the same as those in Fig.$\:$2. As the electron beam traverses the multi-layer films, some positrons between adjacent layers and within the hole get captured, exhibiting a periodic structure; see Fig.$\:$4(a). Relative stability of density distribution of the electron beam, throughout the entire capture process; see Fig.$\:$S14 in \cite{SM}, results in small changes to the self-generated, electrostatic, and transition radiation fields. Therefore, capture and spin evolution of the positrons in each period are similar. Here, we only analyze the detailed capture and spin evolution process of the positrons within the period marked by the blue rectangle in Fig.$\:$4(a). This process can be divided into three stages. Stage$\:$I ($14\:T_{\rm0}\leq t<16\:T_{\rm0}$), positrons move towards the central axis under the action of $E_y^b$; see Figs.$\:$4(c) and 4(e). Meanwhile, $E_x$ begins to form and intensify as the electron beam passes through the sixth layer; see Fig.$\:$4(d). The combined action of $E_x$ and $B_z^b$ causes positrons to move in the $+x$-direction; see Figs.$\:$4(b), 4(d), and 4(f). Under the influence of $B_z^b$, spins of the positrons begin to process: $\overline{S_y}$ changes from $1$ to $0.93$, and $\overline{S_x}$ changes from $0$ to $\pm\:0.3$; see Figs.$\:$4(h) and 4(i). Moreover, due to the relativistic factor $\gamma$ approaches $1$ in this stage; see Fig.$\:$4(g), $\bm{\Omega}_{\rm T}\gg\bm{\Omega}_{\rm a}$, hence the positron spin precession frequency $\bm{\Omega}\approx\bm{\Omega}_{\rm T}\varpropto -B_z\varpropto -B_z^b$; see Figs.$\:$4(f) and 4(j). In Stage$\:$II ($16\:T_{\rm0}\leq t<17.84\:T_{\rm0}$), most positions are located in the hole of the seventh layer at $16\:T_{\rm0}$, and finally pass through the hole driven by $E_x$, moving from $x=15\sim16\:{\rm \mu m}$ to $16\sim17\:{\rm \mu m}$; see Fig.$\:$4(b). The relativistic factor $\gamma$ increases from $1$ to about $2$; see Fig.$\:$4(g). Meanwhile, as the electron beam continues to pass through the seventh layer, $E_\perp$ forms in the hole at $t>17\:T_{\rm0}$; see Figs.$\:$4(e) and 4(f), when the net charge begins to form. Under the action of $E_\perp$, positrons continue to move towards the central axis; see Fig.$\:$4(c). The positron spin procession frequency $\bm{\Omega}\approx\bm{\Omega}_{\rm T}\varpropto -B_z\varpropto -B_{z,{\rm the.}}$; see Figs.$\:$4(f) and 4(j). Consequently, $\overline{S_y}$ decreases from $0.93$ to $0.83$, and $\overline{S_x}$ changes from $\pm$0.3 to $\pm$0.43; see Figs.$\:$4(h) and 4(i). In Stage$\:$III ($t\geq17.84\:T_{\rm0}$), the positrons undergo cascade acceleration in the middle of the adjacent layer and focusing within the hole; see Figs.$\:$4(d), 4(e), and 4(f). As the positron energy increases, the spin precession frequency $\bm{\Omega}$ approaches zero; see Fig.$\:$4(j), so $\overline{S_y}$ and $\overline{S_x}$ undergo no significant changes; see Figs.$\:$4(h) and 4(i). We underline that the multi-layer design allows for capturing more positrons, and depolarization by spin precession is small due to the weak magnetic field experienced by most positrons; see Figs.$\:$4(a), and 4(h).

In conclusion, our novel polarized positron acceleration method combines transition radiation and electrostatic fields to achieve high acceleration gradients, providing a promising solution for accelerator miniaturization and cost-effectiveness. It also addresses challenges commonly encountered in PWFA, such as injection and large depolarization. The high-energy, spin-polarized, and dense positron beam generated through our method has great potential for fundamental physics research and practical applications.

\vspace{1em}

\emph{Acknowledgments:} This work is supported by the National Natural Science Foundation of China (Grants No. U2267204, No. 12275209, No. 12005305, No. U2267204, and No. U2241281), the Foundation of Science and Technology on Plasma Physics Laboratory (No. JCKYS2021212008), the Shaanxi Fundamental Science Research Project for Mathematics and Physics (Grant No. 22JSY014), the Fundamental Research Funds for Central Universities (No. xzy012023046), the Science and Technology Development Fund, Macao SAR (File No. FDCT-0060-2023-RIA1), the Foundation under (Grants No. FY222506000201 and No. FC232412000201), and the Foundation of China Institute of Atomic Energy under (Grant No. YZ222402000401). YIS is supported by an American University of Sharjah Faculty Research Grant (FRG24-E-S29) and acknowledges hospitality at the School of Physics, Xi'an Jiaotong University.

\bibliography{library}

\end{document}